\documentclass[11pt,twoside]{article}
\usepackage{./asp2014}

\aspSuppressVolSlug
\resetcounters

\begin{document}

\title{An Overview of the SKA Science Analysis Pipeline}
\author{C. Hollitt,$^1$ M. Johnston-Hollitt,$^2$ S. Dehghan,$^2$ M. Frean,$^1$
T. Butler-Yeoman.$^1$
\affil{$^1$School of Engineering \& Computer Science, Victoria University of Wellington, PO Box 600, Wellington 6140, New Zealand; \email{Christopher.Hollitt@vuw.ac.nz}}
\affil{$^2$School of Chemical \& Physical Sciences, Victoria University of Wellington, PO Box 600, Wellington 6140, New Zealand}}

\paperauthor{C. Hollitt}{Christopher.Hollitt@vuw.ac.nz}{}{Victoria University of Wellington}{School of Engineering and Computer Science}{Wellington}{Wellington}{6140}{New Zealand}
\paperauthor{M. Johnston-Hollitt}{Melanie.Johnston-Hollitt@vuw.ac.nz}{}{Victoria University of Wellington}{School of Chemical and Physical Sciences}{Wellington}{Wellington}{6140}{New Zealand}
\paperauthor{S. Dehghan}{Siamak.Dehghan@vuw.ac.nz}{}{Victoria University of Wellington}{School of Chemical and Physical Sciences}{Wellington}{Wellington}{6140}{New Zealand}
\paperauthor{M. Frean}{Marcus.Frean@vuw.ac.nz}{}{Victoria University of Wellington}{School of Engineering and Computer Science}{Wellington}{Wellington}{6140}{New Zealand}
\paperauthor{T. Butler-Yeoman}{Tony.Butler-Yeoman@ecs.vuw.ac.nz}{}{Victoria University of
Wellington}{School of Engineering and Computer Science}{Wellington}{Wellington}{6140}{New Zealand}

\begin{abstract}
When completed the Square Kilometre Array (SKA) will feature an unprecedented
rate of image generation. While previous generations of telescopes have relied
on human expertise to extract scientifically interesting information from the
images, the sheer data volume of the data will now make this impractical.
Additionally, the rate at which data are accrued will not allow traditional
imaging products to be stored indefinitely for later inspection meaning there
is a strong imperative to discard uninteresting data in pseudo-real time. Here
we outline components of the SKA science analysis pipeline being developed to
produce a series of data products including continuum images, spectral cubes
and Faraday depth spectra. We discuss a scheme to automatically extract value
from these products and discard scientifically uninteresting data. This
pipeline is thus expected to give both an increase in scientific productivity,
and offers the possibility of reduced data archive size producing a
considerable saving.
\end{abstract}

\section{Introduction}

The volume of data to be generated by a fully operational Square Kilometre
Array (SKA) is staggering. In round figures, the current estimate is that the
SKA will deliver a petabyte of image data per day. It is quite clear that this
data volume is unmanageable with traditional techniques and that as much
analysis as possible will need to be automated.

The science data processor (SDP) of the SKA includes a science analysis
pipeline, which will extract and characterise all scientifically
interesting sources within an observation. The pipeline will present
astronomers with complete catalogues of point, extended and spectral sources
and their properties.

\section{Science Analysis Pipeline Overview}

The science analysis pipeline is conceived as operating in two phases, source
detection and source characterisation. The source detection phase seeks to
exhaustively identify areas of interest amongst the surrounding background in
the first phase. Existing source detection algorithms will play a part in this
extraction \citep{Whiting12, Hancock12}, but will need to be enhanced
\citep{P026_adassxxv} and augmented by extended source detectors
\citep{Frean14, O2-4_adassxxv}.

The source characterisation stage will then extract appropriate science data
products for each detected source. The science products include information
such as spectral index and polarization properties of the sources.

A simplified overview of the source finding and characterisation within the SDP
is shown in figure \ref{fig:architecture}. A initial portion operating at a
relatively high cadence delivers information about the locations of bright
point sources to the sky model used within the pipeline's calibration and
imaging sections. Images and preliminary source information are then passed
into the main data analysis block for further analysis. Where necessary,
there may be provision for users to extract raw images from a buffer at this
stage, even if the SKA does not archive the full raw images.
 
\subsection{High cadence point source finder}
Some initial source finding must be conducted for use in the calibration and
imaging portions of the SKA pipeline. The location and shape information of
bright point sources must be determined in real time and fed back to the
pipeline's sky model.

The source characterisation requirements of the imaging pipeline are somewhat
more relaxed than needed for science purposes, which allows the source finding
algorithms in this portion of the pipeline to trade accuracy for speed.
Nevertheless, the information obtained here is useful as a seed to the later
data processing, as it provides some initial estimates of source properties
that can be refined as necessary.

\subsection{Science analysis pipeline}

At the completion of an observation data enters the main portion of the data
analysis pipeline. Sources are then located and characterised.
While some of the operations required are likely to be computationally
expensive, the relatively infrequency with which they are run (typically every
few hours) means that this will not be a dominant computational driver for the
SDP design. Indeed, much of the processing need not happen synchronously with
observations, but could be conducted at any time before the input buffer is
overwritten by incoming data from subsequent observations.

The first task of the science analysis system is to identify and locate
sources. This is most likely to be performed by a series of three specialised
source finders targeted at point sources, extended sources
\citep{O2-4_adassxxv} and sparse spectral features respectively. These source
finders must be more conservative than existing source finders, so that
valuable data will not be missed.

\begin{figure}
  \includegraphics{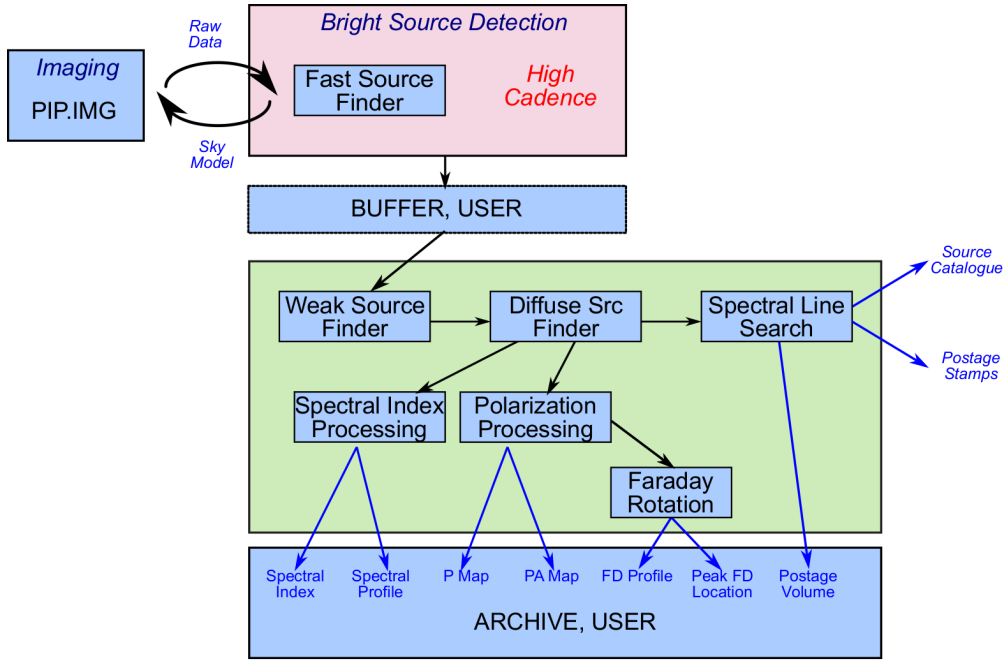}
  \caption{Schematic overview of the PIP.SCI architecture.}
  \label{fig:architecture}
\end{figure}

The lower portion of the architectural diagram in figure \ref{fig:architecture}
shows a nominal set of processes that are used to extract useful data from a
particular observation. In practice a particular experiment might not require
some portions of the processing, but the intent is that the full
characterisation tree would typically be used.

Execution of some elements of the data extraction pipeline will be conditional
on earlier results. For example, full polarization characterisation is only
necessary for polarised sources. Similarly, full spectral profiling of sources
will be necessary only when spectral index fitting fails to adequately
capture the source behaviour.

\section{Discussion}

At least for the foreseeable future, routine storage of the full data set
emerging from the SKA's imaging pipeline would require heroic storage capacity,
having equally unrealistic budget implications. Indeed, full data storage would
render operation of an SKA data archive the dominant operating cost for the
instrument. However, the real time availability of science data products makes
possible considerable savings in data storage. Rather than the commonly assumed
practice of archiving \emph{all} images, one could instead consider storing
only regions where sources were located, while areas of background could be
discarded. For even greater efficiency point sources could be completely
characterised and then discarded from the archive. As astronomical images
typically contain many pixels (or voxels) that contain no useful information,
the potential savings are considerable. Though the choices of trade-off made
have potentially large impact on the reduction of data volume, savings of a
factor of 10 seem straightforward and 100 likely for most observations. Even
higher savings appear possible in special circumstances such as for spectral
line observations.

One could regard selective data retention as a form of scientifically-aware,
lossy data compression. Excision of observed data is currently not a common
operation in radio astronomy, as telescope operating costs are considerably
more expensive than the concomitant data storage. This situation reverses in
the case of the SKA where reobservation will be cheaper than storage.

A careful consideration of whether project resources are better spent on data
collection or data storage appears vital for maximising scientific return from
the instrument. The instinctive desire to store more data comes at the price of
other telescope capabilities. Some segments of the astronomical community are
planning programmes that will have particular demand for data storage. As data
storage costs are expected to decrease with time, it is likely that large
savings can be made by phasing of experimental ambition of storage intensive
projects over time. Science requiring modest storage could be favoured in the
early years of the SKA, while the demanding projects become ever more tractable
as storage costs drop.

The extent to which these ideas will be employed by the early SKA are as yet
unclear. This paper therefore does not represent the formal position of
the SKA project, but simply outlines possible principles of operation
for the science analysis pipeline as conceived by the science analysis
pipeline team of the SKA's Science Data Processor (SDP) consortium.

\section{Unanswered questions}

Specifications for some data products are as yet poorly developed. Should the
SKA move to a science product only storage model, then this becomes a
critical issue. It appears that there has been little engagement of the
scientific community with this matter, to a large extent because of a general
expectation that the SKA will operate similarly to the telescopes of previous
generations. Given the likely operating mode  for the SKA it is vital that the
required data products be identified. The scientific community must therefore
carefully consider the trade-offs when determining requirements.

A second source of uncertainty arises from the unclear boundary between the
SDP and the SKA's data archive. At present the long-term data archive sits
outside the SKA's construction cost cap, and as such details of its operation
remain underdeveloped. Portions of the processing described in
this paper could conceivably be deployed onto the archive or onto regional
science an engineering centres.

\section{Conclusion}
The science analysis pipeline of the SKA seeks to automatically find and
characterise sources within the fields observed by the telescope. It is
unlikely that the SKA project will be furnished with an exhaustive archive, as
is possible for the current generation of radio telescopes. However, some
intelligent winnowing of the incoming data is possible by identifying regions
of scientific interest. Storage of just these regions reduces the storage
requirements of an SKA archive considerably.

\acknowledgements
The authors are  supported in this work by SKA preconstruction funding from the
Ministry of Business, Innovation \& Employment, New Zealand.

\bibliography{O12.7}

\end{document}